# Mission to the Trojan Asteroids: lessons learned during a JPL Planetary Science Summer School mission design exercise

Serina Diniega[a], Kunio M. Sayanagi[b,c,d],Jeffrey Balcerski[e], Bryce Carande[f], Ricardo A Diaz-Silva[g], Abigail A Fraeman[h], Scott D Guzewich[i], Jennifer Hudson[j,k], Amanda L. Nahm[l], Sally Potter-McIntyre[m], Matthew Route[n], Kevin D Urban[o], Soumya Vasisht[p], Bjoern Benneke[q], Stephanie Gil[q], Roberto Livi[r], , Brian Williams[a], Charles J Budney[a], Leslie L Lowes[a]

[a]Jet Propulsion Laboratory, California Institute of Technology, Pasadena, CA, [b]University of California, Los Angeles, CA, [c]California Institute of Technology, Pasadena, CA, [d]Current Affilliation: Hampton University, Hampton, VA, [e]Case Western Reserve University, Cleveland, OH, [f]Arizona State University, Tempe, AZ, [g]University of California, Davis, CA, [h]Washington University in Saint Louis, Saint Louis, MO, [i]Johns Hopkins University, Baltimore, MD, [j]University of Michigan, Ann Arbor, MI, [k]Current affiliation: Western Michigan University, Kalamazoo, MI, [l]University of Texas at El Paso, El Paso, TX, [m]The University of Utah, Salt Lake City, UT, [n]Pennsylvania State University, University Park, PA, [o]Center for Solar-Terrestrial Research, New Jersey Institute of Technology, Newark, NJ, [p]University of Washington, Seattle, WA, [q]Massachusetts Institute of Technology, Cambridge, MA, [r]University of Texas at San Antonio, San Antonio, TX

*Email: serina.diniega@jpl.nasa.gov*

## ABSTRACT

The 2013 Planetary Science Decadal Survey identified a detailed investigation of the Trojan asteroids occupying Jupiter's L4 and L5 Lagrange points as a priority for future NASA missions. Observing these asteroids and measuring their physical characteristics and composition would aid in identification of their source and provide answers about their likely impact history and evolution, thus yielding information about the makeup and dynamics of the early Solar System. We present a conceptual design for a mission to the Jovian Trojan asteroids: the Trojan ASteroid Tour, Exploration, and Rendezvous (TASTER) mission, that is consistent with the NASA New Frontiers candidate mission recommended by the Decadal Survey and the final result of the 2011 NASA-JPL Planetary Science Summer School. Our proposed mission includes visits to two Trojans in the L4 population: a 500 km altitude fly-by of 1999 XS143, followed by a rendezvous with and detailed observations of 911 Agamemnon at orbital altitudes of 1000 - 100 km over a 12 month nominal science data capture period. Our proposed instrument payload - wide- and narrow-angle cameras, a visual and infrared mapping spectrometer, and a neutron/gamma ray spectrometer - would provide unprecedented high-resolution, regional-to-global datasets for the target bodies, yielding fundamental information about the early history and evolution of the Solar System. Although our mission design was completed as part of an academic exercise, this study serves as a useful starting point for future Trojan mission design studies. In particular, we identify and discuss key issues that can make large differences in the complex trade-offs required when designing a mission to the Trojan asteroids.

## 1. INTRODUCTION

NASA's Planetary Science Summer School (PSSS), held annually at the Jet Propulsion Laboratory (JPL), offers graduate students and recent Ph.D. graduates a unique opportunity to design a robotic planetary exploration mission based on the recent NASA Science Mission Directorate Announcement of Opportunity (AO; in this case the 2009 New Frontiers AO (NASA, 2009)). After selecting a recommended mission from the AO, participants formulate a set of science objectives and full mission concept during approximately 10 weekly teleconferences culminating in an intensive week on location at the JPL Team-X concurrent mission design group's facility. This mission concept is then presented to a review panel and thoroughly critiqued. Through this process, participants gain a detailed look at the complex mission design process, systems engineering, and the JPL Team-X concurrent engineering methodology (Wall, 2000; Wall, 2004).

As the participants of the second session of the 2011 NASA PSSS, we elected to design a mission to the Jovian Trojan Asteroids, which has been identified as a priority for future missions by the 2013 Planetary Science Decadal Survey (NRC, 2011). Both NASA (Brown et al., 2010) and the European Space Agency (ESA) (Lamy et al., 2012) have sponsored Trojan mission concept studies.

Our proposed mission would address the questions of how the Trojan asteroids formed and evolved over time by determining the composition, structure, and history of at least two asteroids, assumed to be representative of the population(s) of Trojan asteroids. The answers to these questions have important implications concerning the early history and evolution of the Solar System. We have selected a payload that will accomplish our science mission goals and designed the spacecraft to meet the technical requirements necessary to deliver the instruments to the targeted Trojan asteroids.

This paper will describe the science and engineering aspects of our resultant design: the Trojan ASteroid Tour, Exploration, and Rendezvous (TASTER) mission (Figure 1). Section 2 outlines the science objectives of the TASTER mission. Section 3 describes the selected instrument payload. Section 4 describes the major requirements influencing the design. Section 5 describes the spacecraft itself, including each of the principal subsystems. Section 6 outlines the mission trajectory and schedule for getting to Jupiter's L4 region and the observations that then would be made during the flyby of the first target and the rendezvous with the second (and primary) target. Section 7 describes the risks identified during this study, along with mitigation plans. Cost estimations (calculated in fiscal year 2015 dollars/FY15$, as used in the Decadal Survey (NRC, 2011)) are provided in Section 8.

Finally, Section 9 discusses the main problems that we encountered during this mission design exercise. Although TASTER is the result of a purely academic exercise, our experience in designing the mission within prescribed cost, mass, and power envelopes revealed key issues that future Trojan-targeting mission studies should consider. In particular, we highlight trade-offs and issues involving power and propulsion requirements that resulted from our choice of destinations and mission trajectory. These issues have not been highlighted or adequately discussed within previous mission studies, but we find that careful consideration of a few key points (such as selection of the target Trojans asteroids) at the start of the mission development process should greatly improve the design of future missions. Specific recommendations to that effect are made which we hope will improve the design of candidate missions to these important small bodies.

## 2. SCIENCE GOALS AND REQUIREMENTS

The Jovian Trojan asteroids (henceforth referred to simply as Trojans) are small bodies co-orbiting with Jupiter within the gravitationally stable L4 and L5 Lagrange points. The existence of such bodies was predicted by the French mathematician and astronomer Joseph-Louis Lagrange in 1772 and the first Trojan (588 Achilles) was found in 1906 by the German astronomer Max Wolf. As of 15 Nov 2011, 3175 Trojans have been identified within the leading Lagrange point (the 'Greek camp' and the L4 point) and 1741 have been identified within the trailing Lagrange point (the 'Trojan camp' and the L5 point) in Jupiter's orbit (IAU Minor Planet Center: *http://www.minorplanetcenter.org/iau/lists/JupiterTrojans.html*). It has been suggested that Jupiter's Trojan population may be as numerous as the Main Belt asteroids, with ~$6 \times 10^5$ in the L4 population alone (Yoshida and Nakamura, 2005).

### 2.1 Representatives of small bodies

At the time of this mission study, only 13 small bodies in our Solar System have been observed via spacecraft: 3 bodies have been orbited and 10 have been observed during flyby (Table 1). While these targets span a range of sizes, appearances, and types, bodies 50-200 km in diameter are not well-represented. This size range is thought to contain two important composition transitions: it is within this size range that bodies begin (1) to be more likely to retain physical characteristics set during their accretion epoch and not be changed by collisional processes (i.e., larger than 120 km in diameter in the main asteroid belt (Bottke et al., 2005) and larger than 90 km within the Trojans (Davis et al., 1989; Marzari et al., 1997)) and (2) to undergo thermal differentiation (i.e., larger than 80 km (Hevey and Sanders, 2006)). Structural and compositional studies of our two targeted Trojans would fill this crucial size gap and would greatly aid calibration and refinement of collisional, thermal, and structural evolution models of small bodies. Additionally, a large Trojan is likely to be representative of the thermal, compositional, and radiation environment of its source region within the solar nebula.

### 2.2 Origin Hypotheses and Associated Compositions

There are two leading hypotheses for the source region of the Trojans, and both yield important implications about the Solar System's dynamical history. The first hypothesis is that these bodies formed contemporaneously with Jupiter near their current location and were captured into resonance with Jupiter (Marzari and Scholl, 1998). In this case, these bodies should reflect the solar nebular composition near the ice line and the impactor flux and cratering history within the Jovian system. The second hypothesis is partially motivated by the Nice Model (Tsiganis et al., 2005; Gomes et al., 2005) which predicts that the outer planets migrated inwards within the early Solar System. Due to gravitational perturbations during the planetary migration, bodies within the inner Kuiper Belt would be scattered inwards, with some captured near Jupiter (Morbidelli et al., 2005). In this case, the Trojans should consist of Kuiper Belt material, making these bodies the nearest cache of outer Solar System material.

Evaluation of the general composition (and especially volatile content) of the Trojans would shed light on their likely source region and thereby test these early Solar System evolution models (Rivkin et al., 2010). However, due to their location (within the outer Solar System and co-orbiting with Jupiter), no spacecraft has taken measurements of these bodies. What is known about Trojans comes primarily from ground-based, point-source spectral observations of a fraction of the population. Various studies have compared Trojan spectra to trans-Neptunian objects (TNOs) (Dotto et al., 2008), cometary nuclei (Fernandez et al., 2003; Emery et al., 2006),

Centaurs (Dotto et al., 2008; De Luise et al., 2010), and main belt asteroids (Gradie and Veverka, 1980). Results are inconclusive, however, as spatial resolution of Trojan measurements is generally poor and they have featureless visible to near-IR spectra (e.g., Emery and Brown, 2003; Melita et al., 2008). Evidence for fine-grained silicates has been detected (Emery et al., 2006; Mueller et al., 2010), but no clear volatile signatures have been identified (De Luise et al., 2010). This latter point is most surprising as all source-region and evolution hypotheses predict that the Trojans should be volatile- and organic-rich bodies (Rivkin et al., 2010) and such signatures have been observed on numerous other small bodies (Emery and Brown, 2003), but estimates for surface water content are currently limited to a few percent (Emery and Brown, 2004; Yang et al., 2007). High spatial resolution hyper-spectral images taken by a dedicated spacecraft are needed to more conclusively determine the surface composition of the Trojans (e.g., through the identification of bedrock exposures and tectonic features) and to shed light on the mysterious discrepancies between theory and observation. Detailed spectral studies of the surface of these bodies are also important as that would yield key information about the Trojan's history after accretion, which relates to estimates of the impactor flux and the past and present-day radiative environment within the outer Solar System.

### 2.3 Possible representatives of different source regions

Spectral measurements show a spread in spectral characteristics with two clusters (Figure 2 and Table 2) (Emery et al., 2011). This may indicate that there are at least two compositional populations of Trojans, which may reflect two source regions: Emery et al. (2011) hypothesize that the 'redder' population has comet-like spectra and may have formed in the outskirts of the Solar System, while the 'less-red' population is more spectrally similar to the main asteroid belt and thus may have formed near Jupiter or within the main belt.

Bulk density estimates of two Trojans (that have moonlets) reported in the literature also suggest the possibility of different populations. 617 Patroclus has a very low mean density of 1.08 g $cm^{-3}$, which is likely the result of a very high porosity and significant ice fraction in the interior (Mueller et al., 2010), suggesting formation in the outer part of the Solar System. Despite being of comparable size and albedo, 624 Hektor has a mean density of 2.4 g $cm^{-3}$ (Lacerda and Jewitt, 2007), which is more consistent with main belt asteroids (e.g., Margot and Brown, 2003). However, these density observations contradict the above spectral interpretations as 617 Patroclus is part of the 'less-red' population while 624 Hektor is part of the 'redder' population.

The only conclusion that, thus, can be drawn from this sparse and confusing picture is that observations of all types are needed from more Trojans to determine what characteristics are representative of the Trojan population. It is also vital that a dedicated spacecraft approach and closely observe more than one Trojan (and preferentially from different spectral populations) to provide correlated visual images, compositional data, and gravity measurements that can help determine both the cause for the measured differences and the formative and evolutionary history of a few different Trojans.

### 2.4 Mission science questions and objectives

Based on the science objectives outlined in the Decadal Survey (NRC, 2011) and a survey of current literature regarding scientific analysis, existing knowledge gaps, and the motivations for proposed missions to the Trojans, we discussed and generated a list of top science questions to address with our mission design. These questions relate to two areas:

(1) Where and how did the Trojans form? To address this question, TASTER aims to constrain the origin, compositional heterogeneity, and interior structure of the target asteroids.

(2) How have the Trojans since evolved? To address this question, TASTER aims to determine the geologic and cratering histories, the properties and extent of weathering on the surface, and the presence and composition of complex organic molecules on the target bodies.

Answering these questions would aid NASA strategic goals related to determining how the planets and minor bodies originated in our Solar System and how the Solar System evolved. Our primary science goals are outlined in more detail in our Science Traceability Matrix (Table 3).

## 3. MISSION REQUIREMENTS

In designing a mission, there are two primary classes of requirements that need to be considered in parallel: science and engineering. In particular, the mission payload and spacecraft trajectory must allow the science objectives to be met and the spacecraft and overall mission must remain within the agency and physics prescribed cost, mass, and power envelopes. However, changing one aspect of the mission can influence our ability to satisfy requirements in complex manners. For example, for our spacecraft to reach our target bodies while remaining within power requirements, we needed to consider different propulsion systems for different phases in the spacecraft trajectory. These propulsion options led to different estimates of trajectory time, which in turn affected our estimated mass and mission operations cost. All of these changes influenced our risk estimate with regards to technology development and instrument lifetime. We tackled this constant interplay between different subsystem options using JPL Team X’s concurrent systems engineering approach, which allows subsystems to be designed in real-time while sharing information and parameters with other subsystems (Wall, 2000; Wall, 2004).

A summary of the full science and instrument requirements that we developed is given in Table 3. Our primary science objectives involve determining the geologic, mineral, and volatile composition of at least two bodies, as specified as a requirement within the Decadal Survey (NRC, 2011). Thus, we aim to obtain global visual images of the surface and spectral maps. . In addition, we aim to investigate the interior structure of our rendezvous body through derivation of a detailed shape model and targeted high-resolution images for the identification of local tectonic structures and surface composition variations.

To determine our measurement requirements, we focused only on studies of the rendezvous target and estimated required instrument resolution based on feasible data downlink rates and on the imaging resolution required at various altitudes to achieve our science objectives. In our tour, the flyby target is included primarily for comparison purposes with the rendezvous target, so we aimed for the same resolutions as those required for the rendezvous target body within our surface-based studies (especially regarding observations needed to determine composition, cratering/space weathering processes, and interior/body structure). Thus, the inclusion of the flyby does not alter our instrument requirements; instead, the desired observation resolution was a consideration in the approach distance of the flyby trajectory.

The mission would be a New Frontiers class mission, for which the Decadal Survey recommends a Principal Investigator (PI) managed mission cost cap of $1B (all amounts are given in FY15$), excluding launch costs (NRC, 2011). Our design also took advantage of an optional additional propulsion credit ($17M; adjusted from the $15M in FY09$ presented in the AO using the current NASA New Start Inflation Index) for missions incorporating NASA’s Evolutionary Xenon Thruster (NEXT) ion propulsion system. We included 50% reserve for

Phases A-D and ~25% for Phase E; the Decadal Survey required minimum unencumbered cost reserves of 25% (NRC, 2011). No international cooperation agreements were considered. Additionally, due to the short duration of the PSSS program, we limited our study's scope to the mission design and did not explicitly consider elements that would not significantly affect the full mission concept, such as Education and Public Outreach (EPO) options.

To maximize the deliverable spacecraft mass (i.e., the largest science payload arriving at the target Trojans), we only considered the largest launch vehicle allowed by the AO: a United Launch Alliance Atlas V 551.

Power constraints were driven by spacecraft/propulsion design and payload choice, as the power subsystem needs to provide enough power for both science and trajectory (deep-space and flyby/orbital) maneuvers. In particular, as will be described in the Section 5, power required for the science operations at the rendezvous target (~5 AU) dictated the size of the solar panels. A cruise duration of about 10 years was self-imposed as an initial upper boundary for trajectory selection so convergence to a design could be achieved.

## 4. INSTRUMENTS

Instrument selection for the TASTER mission was driven by our mission science goals (Table 3) and was limited due to strict constraints on the payload mass due to the challenging trajectory design required to reach our target Trojans. We selected a suite of three instruments to accomplish the key mission science objectives within the mass and power limits: WASABI-NACHO -- a dual-camera system, CAVIAR -- a visual and near infrared spectrometer, and ICING -- a neutron and gamma ray spectrometer. Mass and power requirements for all instruments are given in Table 4.

Generally, we aimed to select instruments with extensive flight heritage – this is good mission design practice as it decreases risk in both the development and operational phases; however, we primarily were interested in quickly gaining useful estimates for mass, power, data volume, and cost requirements. Thus, we recognize that our choices may not be optimal with respect to necessary technology development or, equivalently, the ability to utilize heritage, and our instrument suite should be evaluated as a strawman payload. For each instrument, we briefly estimate what changes (if any) from the heritage technology would be necessary. Our proposed TASTER instrument suite, in whole, most closely resembles the payload of the Dawn mission (Russell et al., 2007), which makes sense given the similarities between our missions' science objectives and target bodies.

### 4.1 WASABI-NACHO

The Wide Angle multi-Spectral Asteroidal Body Imager -Narrow Angle Camera Hi-resolution Optics (WASABI-NACHO) instrument is a dual-camera system that is based on the Mercury Dual Imaging System (MDIS) on the MESSENGER spacecraft (Hawkins at al., 2007). The system consists of a multispectral wide-angle camera and a monochrome narrow-angle camera. The data collected would be used to build multispectral visual maps of the Trojan asteroids and perform targeted high-resolution imaging.

The wide-angle camera (WASABI) would have a 10.5°x 10.5° field of view (FOV), with 18 m pixel$^{-1}$ resolution at 100 km altitude. It would have a 10-color filter wheel with one clear filter and two polarizers to provide color imaging over a wide spectral range. The narrow-angle camera (NACHO) would have a 1.5° x 1.5° FOV, with 2.5 m pixel$^{-1}$ resolution at 100 km. These spatial resolutions follow directly from science requirements as outlined in Table 3, with the same FOV as the MDIS. At 100 km, the WAC FOV corresponds to ~5 images to capture a complete hemispherical image of 911 Agamemnon (assuming sphericity).

To evaluate the heritage value of MDIS for our mission, we consider the photon flux the camera would encounter. The solar flux at Mercury is 5.5 x $10^{-3}$ more than at 911 Agamemnon, the bond albedo (~0.12) is higher than that of Trojan asteroids (~0.05-0.08), and the orbital altitudes are roughly comparable between TASTER (100 - 1000 km) and MESSENGER (200 - 15000 km); this means the camera must operate with ~2 x $10^{-3}$ less photon flux at 911 Agamemnon than at Mercury. However, MDIS mostly carried narrow-band filters that were typically 5 nm-wide. For WASABI, we would carry 200-nm wide broadband filters that allow at least 40 times more transmission compared to the narrow band filters of MDIS. With a factor of 40 more flux, the camera would require ~10 times longer exposure to receive sufficient signal. For NACHO, the removal of MDIS NAC's 100-nm medium band filter will increase the photon flux by at least a factor of 5, and the exposure requirement becomes ~100 times longer. The increased exposure times would ensure sufficient SNRs comparable to the MDIS images returned from Mercury and do not pose a problem because of the slower ground-track speed of TASTER compared to that of MESSENGER. MESSENGER's relative ground speed at lowest altitude is 3.3 km/s, whereas TASTER's will only be 15-40 m/s, so smear and the short exposure time would be much smaller concerns at 911 Agamemnon; using exposure times 100 times longer (e.g, 0.1-1 s) at 911 Agamemnon would generate a similarly low along-track smear as MDIS. We believe that these changes to the design and operation of the narrow-angle camera are relatively minor and do not significantly reduce the MDIS heritage.

**4.2 CAVIAR**

The Compositional Analysis from Visible and InfrAred Radiation (CAVIAR) instrument is a "pushbroom" visible and near infrared (VNIR) mapping spectrometer. Similar in design to the Moon Mineralogy Mapper ($M^3$) flown on the Chandrayaan-1 mission (Pieters et al., 2009), CAVIAR would be used to characterize the surface mineralogy of the two target Trojans and to detect the presence of volatile and organic compounds.

With an IFOV of 250 μrad, CAVIAR would have a resolution of 25 m pixel$^{-1}$ at 100 km altitude. This resolution allows the mineralogical composition of discrete geologic units to be directly observed, as well as providing high spatial resolution for determination of surface variability in the presence of water-bearing minerals and organics (Table 3). It would have a spectral range of 0.5-5 μm, which would allow it to detect water and OH features (0.5-3.2 μm; in particular, a fundamental absorption feature at 2.9 μm with overtones at 0.9, 1.4, and 1.9), to detect electronic transition absorptions and vibrational stretching in mineral crystal structures (beyond 0.5 μm), and to measure spectral features up to 5 μm, particularly nitrate and S-O stretch signatures as were measured by OMEGA on Mars Express (Bonello et al., 2004). Additional cooling, beyond that required for $M^3$, would be needed to enhance the SNRs due to the extension of the spectral measurements into the mid-wavelength infrared, but this should not require a significant change in design as the mission targets are also further from the sun. Like $M^3$, the instrument will detect 260 bands at full spectral resolution, yielding a spectral-resolution of ~20 nm; although this is half the spectral-resolution of $M^3$ (Pieters et al., 2009), it is of similar resolution as OMEGA within the mid-wavelength infrared (Bonello et al., 2004) and would be sufficient for detection of many of the spectral signatures of interest as long as SNR is sufficiently high (> 100, based on design specifications of OMEGA (Bonello et al., 2004)). $M^3$

was designed for an SNR of > 400 when taken measurements in the lunar equatorial region (Pieters et al., 2009), so the heritage value of $M^3$ should be retained.

**4.3 ICING**

The Instrument for Collection of Incident Neutrons and Gamma-rays (ICING) would map the near-surface (to a depth of 1 m) abundances of volatile compounds and ices (H, C, N and O) in the upper-meter of the rendezvous target and unambiguously identify and map the major rock-forming elements (O, Si, Ti, Al, Fe, Ca, and Mg). Operated continuously during orbit of the rendezvous asteroid, ICING would record the spectra of gamma-ray and neutron energy emitted from the surface regions under the spacecraft. The integrated signal would build a full-coverage map of the surface with resolution of 80° or less. This essentially allows hemispherical-scale gross compositional and volatile abundance variability to be assessed, which addresses scientific questions of 911 Agamemnon's compositional heterogeneity and formation history (Table 3).

The design of ICING is directly adopted from the GRaND instrument on the Dawn spacecraft (Prettyman et al., 2003; 2011; Russell et al., 2007) and draws on experience from the successful Mars Odyssey and Lunar Prospector missions. Minimal changes should be needed for adopting this instrument for use as epithermal neutrons are predominantly produced by galactic cosmic rays (Prettyman et al., 2006) whose flux is independent of the distance from the sun.

**4.4 De-Scoped Instruments**

Several additional instruments were considered for the TASTER mission and de-scoped during the mission concept design process (Table 5). Mass constraints limited the proposed spacecraft to only the instruments that could collect the highest-priority scientific data; less-critical data collection objectives and duplicate functionality with the selected instruments caused these instruments to be assigned a lower-priority and eventually eliminated from the proposed TASTER spacecraft design.

A radar ranger/sounder was considered, but ultimately eliminated from the proposed instrument payload due to its high estimated mass (17 kg) and because some structural data about near-surface features of the Trojan could also be determined from measurements by the ICING instrument. A thermal/IR imager was de-scoped because CAVIAR could take similar measurements. Likewise, a laser altimeter was de-scoped because similar topographical mapping and feature identification could be accomplished with stereo-imagery collected by WASABI-NACHO. An ultraviolet spectrometer was eliminated because it overlapped in capability with ICING, the likelihood of detecting outgassing from a Trojan asteroid was estimated to be small, and because the light forward-scattered by outgassed particles could be detected by WASABI-NACHO during solar occultation. Finally, the option of an impactor during the flyby was eliminated due to its attendant increase in mission complexity.

**5. SPACECRAFT DESCRIPTION**

The TASTER spacecraft design (Figure 3) follows guidance from the 2013 Planetary Science Decadal Survey (NRC, 2011) and the NASA New Frontiers AO (NASA, 2009).

**5.1 Spacecraft Overview and Configuration**

The main spacecraft bus is a 2x2x2 m cube, with a best mass estimate of 275.9 kg. The spacecraft was designed to fit inside an Atlas V-551 short fairing for launch and features two folding-wing solar arrays, with the science instruments mounted on the nadir-facing (target-facing) side (Figure 3). A fixed high-gain antenna, the hybrid propulsion system (consisting of a dual propellant chemical system and twin NEXT ion engines, see Section 5.2) and solar wing mounts occupy four of the remaining five sides of the spacecraft bus with the zenith-facing side empty. The long solar panels allow the spacecraft to have a relatively small moment of inertia along the y-axis, which allows the spacecraft to efficiently and quickly slew during the flyby phase. Given these design considerations, the spacecraft is 3-axis stabilized using reaction wheels and thrusters (see Section 5.4). The main propulsion engines are operated on the anti-ram side (z-axis) of the spacecraft, with the NEXT engines attached to the spacecraft with gimbals.

Incorporating JPL Design Principles Margin of 30%, the spacecraft's mass is 1187 kg (1998 kg including propellant). The primary mass drivers are fuel (811 kg -- required for the nearly 4 km $s^{-1}$ of delta-V necessary to reach the target Trojans) and the solar power system (396 kg, which includes large structural supports for the solar arrays: 122 kg, and large arrays: 57 $m^2$ -- required to provide sufficient power at ~5 AU (see Section 5.3)). The chemical fuel (70% of the propellant) would be accommodated in an efficient volume by using four propulsion tanks (two hydrazine ($N_2H_4$) and two NTO (di-nitrogen tetroxide: $N_2O_4$)) in the lower corners of the spacecraft bus with the Xenon gas tank for the ion propulsion in the center.

Spacecraft design was driven by a requirement for high reliability over an 11-year mission duration and the use of commercial off-the-shelf (COTS) parts, when available. Systems are dual-string redundant with the second string off during normal operations. In addition, we would extend the design phase to allow for additional system design work. Radiation dose to the spacecraft would be minimal during the mission: 28.2 mrad behind 100 mils of aluminum with the bulk of the dose experienced during the Jupiter flyby (at an altitude equal to 35 Jupiter radii). No new technology would be required for the spacecraft design and NASA would be responsible for bringing the NEXT engines to Technology Readiness Level (TRL) 6 per the New Frontiers AO (NASA, 2009).

No science instruments are deployable, which reduces complexity and risk. Thermal vacuum and vibration testing would be conducted at JPL and assembly, test and launch operations would be conducted at Kennedy Space Center, FL (KSC).

## 5.2 Propulsion

The spacecraft trajectory (described in Section 6) would be accomplished through the use of a combination of traditional chemical and next-generation ion engines. The selection of NASA's NEXT ion engine diverges somewhat from the philosophy of using exclusively COTS technology, but the mission profile benefits significantly from its inclusion as it increases mission capability while remaining within mass and power requirements (as was demonstrated by the Dawn mission (Rayman et al., 2007)).

According to the 2008 development status update (Patterson and Benson, 2007), the NEXT ion engine represents a significant improvement over the previous generation of technology. Each NEXT unit is capable of producing 0.24 N of thrust, with a specific impulse $I_{sp}$ of 4100 s, delta-V of 2779 m $s^{-1}$, and a maximum power requirement of 7 kW.

Aerojet's 100 lbf High Performance Apogee Thruster (HiPAT) dual-mode chemical engine was selected for its ready availability and reliable history of operation. This motor utilizes a hydrazine and NTO mixture in bi-propellant mode, providing a maximum of 445 N with an $I_{sp}$ of 328 s and a delta-V of 1135 m $s^{-1}$.

## 5.3 Power

Solar power was chosen for this mission concept to minimize risk, which resulted in an increase in mass (as discussed in Section 5.1). Given the limited quantity of Plutonioum-238 (Pu-238) available for radioisotope thermoelectric generators (RTG) in the next decade (DOE, 2010), Advanced Stirling Radioisotope Generators (ASRG) would be the only feasible RTG since they operate with significantly less Pu-238 than the Multi-Mission RTGs (MMRTG). However, as ASRGs have yet to operate successfully on a long-duration space mission, this was seen as a riskier option given the 11-year primary TASTER mission duration. To mitigate risk, backup ASRGs could have been added to the spacecraft, but this significantly increased the mass of the spacecraft.

Our solar power system was designed to accommodate two peak situations: science operations requirements at ~5 AU and the power requirements for the Solar Electric Propulsion (SEP) maneuvers; the latter was the larger requirement and thus dictated the required solar panel surface area. Our mission design operates the SEP system to a maximum solar distance of 3.9 AU, at which point the spacecraft requires 1428 W of power (including 30% contingency). To accomplish this, 57 $m^2$ of rigid Gallium-Arsenide (GaAs) triple-junction cells operating at 30% efficiency are needed – yielding a design comparable in area and efficiency to the solar panels currently operating on the Juno spacecraft (Grammier, 2009), but still a technological challenge (as will be discussed in Section 9). The panels would be 2-axis articulated to track the Sun and could produce 23 kW at 1 AU, 1530 W at 3.9 AU and 930 W at 5 AU.

We note that Brown et al. (2010) had estimated that >300 $m^2$ of solar panels were required to operate a SEP system beyond 3.5 AU. Although not explicitly stated by Brown et al. (2010), it seems that this much higher solar array area estimate assumes a low solar cell efficiency (while we assume Juno-quality solar cells) and that solar power would be used for the rendezvous maneuver (we use chemical propulsion; Brown et al. (2010) briefly discusses a mission architecture that uses a similar hybrid propulsion system and there estimates a more comparable solar array area of 86 $m^2$).

Our battery requirements were dictated by the length of time from launch through solar panel deployment: the spacecraft would require 422 W, including 30% margin, during launch operations with all power coming from three lithium-ion batteries (two primary and one backup). In addition to launch operations, the batteries would be in use during the flyby phase to provide additional power during this short-peak in science and telecommunication operation. The batteries have a 24 A-hr capacity and sufficient power to satisfy the spacecraft and payload needs even after a 70% depth-of-discharge.

### 5.4 Attitude Control System

The Attitude Control System (ACS) concept would have sensing capabilities provided by two precision star trackers, two inertial measurement units, and one internally redundant sun sensor assembly for attitude determination during most phases of the mission. Attitude control is achieved by a set of monopropellant ACS low-thrust engines, the gimbaled NEXT engines (during the electric cruise phase), and a pyramid assembly of 4 reaction wheels for normal pointing. Each reaction wheel has 12 N m s angular momentum and 0.075 Nm torque capability for fine pointing requirements. The ACS is designed to provide three-axis stabilization in order to enable science data measurements and calibrations.

The driving requirement for ACS design is the NACHO bore-sight pointing requirements of 206 arcsec (3-sigma) with 90 arcsec (3-sigma) for attitude determination of both the instrument and the High Gain Antenna (HGA). We also require that the jitter be kept below 10 arcsec/axis for a stable control system, which satisfies the NACHO stability requirement of 50 arcsec/axis. This system can provide a slew rate of up to 0.3° $s^{-1}$ during flyby. The angular rates required for orbiting altitudes of 1000 km, 500 km, and 100 km orbits about the rendezvous target are 0.001, 0.004, and 0.015° $s^{-1}$, respectively.

The ACS design also ignores external torques, i.e. it neglects a gravity gradient. It assumes that no magnetic torque would be induced by the target body and neglects the solar pressure at 5 AU (dependent upon the surface area of the spacecraft).

**5.5 Data and Software**

The Command and Data Systems (CDS) would provide all of the computational needs for the mission, including control and data processing for the ACS, instruments, power, and communication of science data and spacecraft operational status to Earth. Dual RAD750 processors are fully redundant and are in a "cold" dual string configuration, where one processor is on and functioning, while the other starts the mission powered down but turns on if a fault arises. Each processor runs at a clockspeed of up to 200 MHz and has level 1 and level 2 cache sizes of 32 KB and 256 KB, respectively. Processing and bus margins are large enough to allow all subsystems to operate simultaneously.

Instrument-specific interface cards connect the science instruments to the CDS processing core utilizing the Multi-Mission System Architectural Platform (MSAP). To handle the high data throughput from CAVIAR, interfacing with the processing core is done with a non-volatile memory and camera (NVMCam) card that contains 32 Gb of memory for data storage. WASABI-NACHO is also connected to the core through this interface. ICING is connected to the core through a MSAP System Interface Assembly (MSIA) interface that provides fault detection and a connection to the backup processing string. A data record rate of 20 Mbps and a playback rate of 13 kbps are anticipated; these rates are determined by the telecommunications downlink properties via the MSAP Telecommunication Interface (MTIF) card. The MTIF card also provides interfaces to the power subsystem and the majority of ACS guidance systems.

The software for the SEP interface to the ACS, command and data handling, fault protection, power, system services, telecommunications, and thermal systems are inherited from Dawn (Russell et al., 2007). Similarly, individual instrumental software packages are also inherited from their predecessors (WASABI-NACHO from Messenger MDIS, CAVIAR from Chandrayaan-1 $M^3$, and ICING from Dawn GRaND). The considerable inheritance of these tested and successfully deployed software components would greatly lower software development costs and mitigate mission risk.

**5.6 Thermal**

The thermal system aims to keep the spacecraft temperature between 240-325 K for proper instrument and electronics operation; this would require passive cooling at 1 AU (hot phase) and active heating at ~5 AU (cold phase). The spacecraft bus is surrounded with multi-layer thermal insulation and have two heaters designated for each of the three science instruments. A 1.3 $m^2$ louvered radiator helps regulate internal temperature by opening the louvers, radiating heat to space, in the hot phase and closing the louvers in the cold phase. During the cold phase, the heaters require 20 W of power to maintain the internal temperature.

**5.7 Communications**

The telecommunications subsystem would provide for communication between the spacecraft and the Operations team. It uses the Deep Space Network (DSN) X-band (8-12 GHz) transceivers, which allows for the simultaneous uplink and downlink of commands, science data, spacecraft status, and telemetry. The maximum allowable uplink/downlink data rate is approximately 9.5 kbps, so an 8-hour per day downlink session would yield a total data transmission volume of 273 Mb per day.

The spacecraft has one fixed 1.75m high gain antenna (HGA) mounted along the +z axis of the spacecraft, which allows for maximum data transmission. One medium gain antenna (MGA) and 2 low gain antennae (LGA) are located on the +x, -x, and –z panels of the spacecraft, respectively (Figure 3). There are also two small deep space transponders (SDSTs) and two 25W travelling wave tube amplifiers (TWTAs) located on the +y panel. The SDSTs have been adopted from the Dawn and Mars Reconnaissance Orbiter missions. Signals received by the HGA, MGA, or LGA antennae are passed to diplexers, which provide transmission and reception capabilities, then to the TWTAs, which provide signal amplification. The signal then proceeds to the SDSTs, which demodulate the signal into ranging and command components, and transfer the command components to the CDS subsystem. For transmission, science and telemetry data are encoded by CDS, modulated by a ranging component stored in the SDSTs, amplified by the TWTAs, and transmitted by the appropriate antenna given the circumstances. The system is fully cross-strapped, allowing for the distribution of redundant data to the SDSTs.

Reception of downlinked data is accomplished via the three sites of the DSN: Canberra, Australia, Madrid, Spain, and Goldstone, California, USA. Each site consists of one 70 m station accompanied by a group of 34 m beam waveguide antennae (BWG). The mission concept for the ground systems portion has also been adapted from the Dawn mission (Russell et al., 2007).

## 6. Mission Design

For our proposed mission, Phase A would commence in 2013 and last for 12 months. Phase B was extended to 15 months to allow for additional design time (as mentioned in Section 5.1). Phases C/D would commence in 2015, yielding 40 months for manufacture and testing. The initial launch window extends from January 20 to February 9, 2019, with a 21-day launch window opening every 13 months for trajectories to the Trojans (although the specific target selections may not be repeatable).

Our selected Atlas V launch vehicle in a 551 configuration enables a spacecraft total mass launch capability of 1974 kg with a characteristic energy (C3: the square of the hyperbolic excess velocity at which the spacecraft departs Earth) of 53.1 $km^2 \cdot s^{-2}$. We note that while this value is lower than that required by Brown et al. (2010) ($C3 \geq 73.5\ km^2 \cdot s^{-2}$), this C3 is sufficient for our mission design as we include SEP-provided propulsion during the cruise stage. This provides 2.4 times more delta-V during cruise, thus decreasing the amount of energy required at launch.

### 6.1 Trajectory and Target Choice

The mission flight profile is optimized for an 11-year operational life, with one primary rendezvous target and one flyby target selected from the group of Trojans which librate stably in Jupiter's L4 Point. In order to accomplish the mission's science objectives, the large and relatively well-studied 911 Agamemnon was chosen as the primary target; the flyby target, 1999 XS143, was then selected from those bodies requiring the least amount of course change during the cruise phase. We note that the flyby target was selected somewhat arbitrarily as none of the candidate bodies have been studied and all are small and/or dark. As will be discussed in Section 9, we recommend that a more thorough survey of possible rendezvous targets be considered and that promising flyby targets be targeted by Earth-observations and characterized, prior to final target selection.

The cruise phase would be characterized by a long duration flight with two deep space prolonged thrusting periods by the SEP system. A Jupiter gravity assist occurring between the two SEP burns would change the trajectory inclination to match that of the orbital plane of our target Trojans (~22° above the ecliptic). The total delta-V required is 3.914 km $s^{-1}$ which includes the chemical engine burn required to match Agamemnon's velocity at arrival. The power demands from the science instrument payload are minimal during the cruise phase, thus allowing power generated by the solar panels to be primarily used to drive the dual NEXT ion engines. Although the operation of both thrusters are required to reach the target bodies via our planned trajectory (Figure 4), the redundancy would allow for mission continuation in the case of a single engine failure, with a modified flight trajectory and extended timeline.

### 6.2 Flyby of 1999 XS143

The spacecraft would approach the asteroid 1999 XS143 to within 500 km altitude, which was the minimum altitude judged to be safe against possible collision with a moonlet (i.e., outside the Hill Sphere), at a flyby speed of 2.56 km $s^{-1}$. NACHO would begin taking images as soon as the target is resolvable (i.e., the diameter of 1999 XS143 will cover 5 pixels ~100 hours before closest approach or ~$10^6$ km distance) to collect information about the surrounding environment for detailed navigation corrections and to begin science observations. For the last 5 hours of approach and the first 5 hours of departure, WASABI and CAVIAR would also be operational, collecting additional visual and spectral images. No propulsion burn operations other than attitude control would occur during this time.

### 6.3 Orbits of 911 Agamemnon

Seven months after the flyby of 1999 XS143, the spacecraft would enter a matching orbit of 911 Agamemnon and achieve gravity capture. This maneuver requires a considerable delta-V (1135 m $s^{-1}$) that would be supplied by the chemical engine, Aerojet HiPAT 445N (unused until this point in the mission), to rapidly slow the spacecraft and effect a stepped orbital decay.

Upon orbital insertion, science operations are to be conducted at three discrete orbital altitudes within a few different polar orbits (Table 6). This tiered-approach was selected to mitigate navigational risk and to improve science observation planning; a similar method has been successfully employed by Dawn in approaching and observing the main belt asteroids Vesta and Ceres (Russell et al., 2007), where observations and the shape model generated during one orbit can be used to alter and refine navigation and science planning for the next (closer) orbit.

In the first phase, at 1000 km from the surface, WASABI-NACHO and CAVIAR will begin a global imaging campaign; these images will begin science operations and allow generation of a detailed shape model, gravity map, and rotation measurement. This phase would last 20 days, with an orbital period of 120 hours, and relative ground speed of 15 m $s^{-1}$.

The second phase, at 300 km from the surface, with a 24-hour orbit and 30 m s-$^{1}$ relative ground speed, would last 3 months. Higher-resolution visual images and gravity measurements will be recorded, allowing for refinement of the target's shape model and planned future science observations. During this phase, ICING would begin operation.

The lowest orbit would be 100 km above the surface, with an 8-hour orbit and relative ground speed of 40 m $s^{-1}$. The spacecraft would switch between two polar orbits after 6 months, to yield different illumination/viewing angles. In this phase, all of the instruments would continue to be active, with WASABI-NACHO performing targeted high-resolution imaging, and ICING completing a global particle energy map with a resolution of 40°. After eight months on this orbit, main fuel reserves are expected to be exhausted and the nominal mission would end. A possible mission extension would then be possible due to the flexibility given by the ion propulsion system, depending on the amount of residual Xenon propellant. If electrical power was not used by the instruments it would be possible to throttle the NEXT engine to lower values and perform an orbital maneuver to further destinations. The feasibility of this type of maneuver is currently being demonstrated by the Dawn mission as this spacecraft is departing from its rendezvous with Vesta and will move towards to a rendezvous with Ceres in 2015 (Russell et al., 2007).

## 7. RISK ASSESSMENT

Primary mission and programmatic risks were identified throughout the concurrent mission design process, and were considered in trade-off analysis. Based on impact and likelihood, three main risks within our final mission design are discussed, along with their corresponding mitigation strategies.

(1) There is a severe scarcity of data for all Trojans – most Trojans lack even lightcurve observations and thus their size and rotation rate are unknown, which makes it difficult to plan flyby or rendezvous approaches. The proposed mitigation consists in choosing our rendezvous target from the small set of bodies with more detailed prior observations and in rendezvousing at several (decreasing) altitudes, so that adjustments during the mission can be made to planned navigation and science operations. Additionally, a ground-based survey program early in the mission development cycle can help acquire needed basic information about potential rendezvous and flyby targets -- such as size, rotation rate, and presence of a moonlet. After launch, pre-encounter imaging will fill knowledge gaps and aid in making decisions about the optimal flyby altitude. For the rendezvous target, sequentially decreasing orbital altitudes, each maintained for the week(s) needed for proper data analysis to be incorporated into the flight plan, can be used to assess risk and plan the next mission phase (as was done for Dawn's approach and orbit of Vesta (Russell et al., 2007)).

(2) The prime mission duration is 11 years, including 10 years of cruise. This turns the reliability of components and their lifetimes into a risk requiring proper operation strategies and redundancy mitigation. To ensure that critical subsystems can survive and perform during the long mission duration, failure modes and effects must be properly identified. The proposed mitigation includes performing Parameter Trend Analysis (PTA) on instrumentation with mean life expectancy less than five mission lifetimes. A testing plan that accounts for parts up-screening and burn-in runs is considered, along with dual-string redundancy for crucial components. In addition, events such as trajectory corrections and power-cycling of subsystems during SEP cruise phases were identified as mission critical. Mitigation strategies benefit by inheriting procedures and know-how by JPL's multi-mission operations system on previous ion propulsion-powered missions such as Dawn (Rayman et al., 2006).

(3) The hybrid propulsion system considered for TASTER makes use of the NEXT ion engine, which is currently at a TRL below 6. This is a NASA-managed program component (Patterson and Benson, 2007), which means that the cost and schedule risk of technology development falls to NASA. However, our mission design still has a large schedule risk if those engines are not tested and integrated in time. To mitigate this risk, a schedule extension fallback plan should be incorporated into the mission design.

## 8. COST

The cost estimate for our mission was $1005.5M, which was below the cost cap of $1017.4M ($1B AO cap + $17M propulsion credit in FY15$; where necessary, values have been adjusted based on the NASA New Start Inflation Index listed during August 2011). This amount was estimated using the JPL Team-X methodology, which uses a mix of quasi-parametric and grassroots algorithms. Table 7 shows the breakdown of cost per mission phase, along with reserve (50% for most phases; the reserve amount is lower for Phase E as we did not include reserve on tracking costs). The primary cost driver during the development costs (Phases A-D) would be the flight system; its cost came to $332.9M (out of $792.2M), with the largest expenses in Power ($49.7M), Structures ($53.2M) and Propulsion ($81.9M) due to our trajectory choice, which dictated our delta-V and power requirements. Our trajectory choice also dominated the operations costs; mission operations would be the largest Phase E cost at $72.9M (out of $183.3M total), due to the long cruise-time and required monitoring of the NEXT ion propulsion engine.

De-scopes are possible to lower the cost, such as cutting 6 months of observations from the rendezvous phase which would save approximately $20M. This decision would not need to be made until late in the development phases (Phase E), so this could mitigate small late-stage cost-overrun issues that go beyond our included cost reserve.

## 9. MAIN LESSONS LEARNED

Most of the major design issues, large risk obstacles, and power, propulsion, and cost difficulties that we encountered during this mission design exercise related to mission constraints dictated by our target and trajectory choice. Thus, these problems can be partially mitigated through a more careful target and trajectory selection. Here, we will briefly describe the main issues and then make suggestions for future mission concept designs for spacecraft visiting the Trojans.

### 9.1 Power Issues

Our review panel expressed a large concern about the decision to use solar-power as they thought that it would be challenging to engineer panels and its associated array drive systems to withstand the transient load imparted to the spacecraft at orbital insertion by the HiPAT bi-propellant engine. However, it was considered a reasonable solution by our Team-X experts and a prior Team-X study (Bonfiglio et al., 2005) also concluded that solar-power should be feasible for powering NEXT ion engines and scientific instruments during a Trojan mission. Additionally, as previously mentioned, the similar array size of the Juno mission provides some technological heritage in order to tackle this problem. Juno will perform a similar orbital insertion burn (30 minutes) for a bi-propellant engine rated at 640 N (AMPAC-ISP Corporation LEROS 1B).

*Recommendation*: A mission to the Trojans has a number of unique constraints, such as distance from the Sun, and requirements, such as high amounts of delta-V needed to increase orbit inclination that make it supremely difficult to remain within the New Frontiers class cost cap while delivering a reasonable science payload. This is naturally further complicated when the mission trajectory needs to include more than one target. We found that hybrid propulsion and power solutions were needed and thus recommend this approach. Although this expands the trade-space beyond the traditional chemical vs. solar vs. nuclear studies, the consideration of combinations should increase allowable payload mass (generally through a tradeoff with structure requirements and trajectory length). We also highlight that radioisotope power sources can be considered in place of (or in addition to small) solar panels (Bonfiglio et al., 2005 and Brown et al., 2010). (In our study, due to time constraints, an ASRG power source was considered but quickly discarded due to issues pointed out in Section 5.3, and per Section 5.2.5.2 of the NF AO (NASA 2009).) Additionally, as the JUNO mission demonstrates (Grammier, 2009), recent improvements in solar panel efficiency make solar power feasible even out at Jupiter's orbit without requiring unreasonably large arrays (as had been concluded by Brown et al., 2010).

Non-direct trajectories should also be considered in the tradeoff study: additional gravity-assists from Earth, Venus, and Jupiter can be added to decrease the amount of delta-V required (which translates into less propellant/mass) (Simanjuntak et al., 2010), although this generally will increase mission (cruise phase) duration and associated operations costs (Bonfiglio et al., 2005) and associated risks.

**9.2 Propulsion Issues**

Due to the large delta-V requirements for a mission to the Trojans, use of an advanced propulsion system is needed to deliver a reasonable amount of mass while remaining within the total mass, power, and cost requirements of the New Frontier mission class (or Discovery mission class (Perozzi et al., 2001; Rayman et al., 2007)). A hybrid propulsion system was chosen which combines an efficient ion engine cruise and a quick delta-V chemical engine. The use of this type of system would provide more delta-V during the cruise stage and decrease the launch energy required (C3), thus allowing a heavier spacecraft to be flown. However, due to the NEXT ion propulsion system's low thrust, a longer mission cruise time would be required to fully exploit this engine (Rayman et al., 2007) and this creates an increase in mission complexity/operations costs. Additionally, as discussed in Section 7, development of this engine is a NASA-managed program component, which alleviates cost concerns but would create a large schedule risk.

Due to the long time needed to build thrust, problems were also encountered when balancing science observations with time required for engine activity – in particular, during approach to the target bodies. To decrease the overlap in time requirements, we chose to use chemical propulsion during orbit insertion for the primary body.

*Recommendation:* Careful analysis should be done of propulsion possibilities, including possibilities that are not yet at TRL 6, and tradeoff studies should consider cruise time as well as cost, power, and mass requirements (Rayman et al., 2007). Hybrid systems that utilize both chemical and advanced systems, such as those proposed in this study, should also be considered, especially for maneuvers that would be concurrent with science observations, such as during orbital insertion. Fortunately, solar electric propulsion and other advanced propulsion systems are being used in a number of current and upcoming missions, so development and demonstration of relevant technology options is ongoing. For example, Dawn uses an ion propulsion system for all post-launch trajectory control and corrections, including rendezvous and orbit insertions (Rayman et al., 2007) and BepiColombo, to be launched in 2015, will use a standard chemical propulsion system for Earth escape and Mercury orbit insertion and a solar electric propulsion system for the rest of the interplanetary cruise phase (Schulz and Benkhoff, 2006).

### 9.3 Trajectory and Target Selection

The importance of target and trajectory selection during an early mission design phase for the optimization of the mission cannot be over-emphasized. With the TASTER mission design exercise, we found that the target and trajectory selection was a key driving factor (especially in remaining within the mass and power envelope) and thus dictated many of the choices during tradeoff studies. Due to the overall lack of data for all but a handful of Trojans and study time-restrictions for assessing how target selection affected trajectory design and delta-V requirements, it was not possible for us to perform a detailed tradeoff study among target possibilities. We instead calculated rendezvous trajectories for the six "oldest" Trojans (as generally larger Trojans were observed earlier and more often): 588 Achilles (1906 TG), 617 Patroclus (1906 VY), 624 Hektor (1907 XM), 659 Nestor (1908 CS), 884 Priamus (1917 CQ) and 911 Agamemnon (1919 FD). The amount of information about Agamemnon and the relatively low required delta-V were our primary reasons for selecting this as our rendezvous target. Five possible flyby targets were then examined, based on a search for bodies that would be close to the direct trajectory (2002 EO144, 1999 XN226, 1995 QD6, 1986 TT6 and 1999 XS143) and one was chosen based on delta-V requirements and estimated size.

*Recommendation*: As discussed in Perozzi et al. (2001) for identification of rendezvous and flyby Near Earth Asteroid targets, one must consider targets with a view towards satisfying *both* scientific and engineering requirements. Thus, we suggest that future studies invest in an extended study of possible rendezvous and flyby targets, of which the primary metrics are (1) science value, (2) delta-V requirements, and (3) mission duration.

This study for the identification of possible targets must be done in the early phases of mission planning as the location (~5AU) and orbital characteristics of the Trojans (especially their inclination) control many decisions about power, propulsion, and spacecraft structure. The primary focus should be on the rendezvous target; although current observations indicate it is necessary to visit at least two bodies (of different types) to adequately characterize the full Trojan population. The ESA study in fact recommends a mission to at least 5 Trojans (Lamy et al., 2012). Fortunately, clustering of the Trojans makes it very likely that at least a dual Trojan encounter will be possible (Bender and Penzo, 1995).

In starting the list of possible rendezvous targets, the first objective is to limit the energy required to reach the target. Generally, this corresponds to the body having a low inclination as out-of-plane maneuvers are very demanding in energy requirements (Perrozi et al., 2001). Low eccentricity is also desirable to minimize the delta-V and propellant required. To simplify mission planning and maximize scientific return, the Trojan should also be well-studied and large enough to be representative of the original accretion population (i.e., not a collision fragment).

Thus, an initial list of possible rendezvous targets can be found by surveying the literature and identifying Trojans that have previously been studied with (1) low inclination, (2) low eccentricity, and (3) large diameter/small absolute magnitude. In Table 8, we present a starting list of possible rendezvous targets compiled from several prior Trojan mission design studies and reports of ground-based telescope observations. We do not claim that this list is comprehensive, but it is more thorough than lists considered in prior mission studies (including ours) and thus should provide an advanced starting point for future mission studies.

Once a few possible rendezvous targets have been chosen as most viable, potential flyby targets can be selected based on proximity to the initial trajectories. When these bodies have been identified, a formal observation campaign should be undertaken to determine light-curves and spectra for unstudied bodies. Fortunately, simple Earth-based spectral observations are sufficient to at least identify the spectral class of potential targets, either redder or less-red (Emery et al., 2011). The final rendezvous and flyby target(s) can then be chosen based on the flyby-added energy and cruise time and overall science value.

Finally, we note that some prior mission studies proposed also observing (generally via a flyby) a Main Belt Asteroid during the cruise phase out to the Trojans (Lamy et al., 2012). This option was not considered in our mission design, but its inclusion would certainly increase the science value of the mission and may not significantly increase the required mass, power, or cost if the trajectory were designed with this in mind during early mission planning.

## 10. Conclusion

Based on our participation in the NASA-JPL Planetary Science Summer School and our mission design of TASTER, we gained valuable experience in the art of designing a mission within prescribed cost, mass, and power envelopes. In particular, we learned of key trade-offs and issues involving power and propulsion requirements that resulted from our choice of destinations and mission trajectory. In presenting our mission and the lessons we learned, we aim to convince future Trojan mission design studies to carefully consider a few key points -- such as careful selection of the target Trojans -- at the start of the mission development process. We hope that these suggestions will aid future efforts to visit these intriguing small bodies as a dedicated mission appears necessary if we are to unveil the formative and evolutionary history of the Trojans and feast on a wealth of unique information about the early Solar System.

## Tables

| Name | Body-type | Observation type | Mission/year of visit | Diameter |
|---|---|---|---|---|
| *Ceres* | *Main belt asteroid/dwarf planet* | *Orbit (3 altitudes)* | *Dawn/2015* | *950 km* |
| *4 Vesta* | *Main belt asteroid* | *Orbit (3 altitudes)* | *Dawn/2011-12* | *530 km* |
| ***911 Agamemnon*** | ***Trojan*** | ***Orbit (3 altitudes)*** | ***TASTER*** | ***167 km*** |
| 21 Lutetia | Main belt asteroid | Flyby (primary: 67P/Churyumov-Gerasimenko | Rosetta/2010 | 100 km |
| **1999 XS143** | **Trojan** | **Flyby (primary: Agamemnon)** | **TASTER** | **60 km*** |
| 253 Mathilde | Main belt asteroid | Flyby (primary: Eros) | NEAR Shoemaker/1997 | 50 km |
| 243 Ida | Main belt asteroid | Flyby (primary: Jupiter) | Galileo/1993 | 30 km |
| *433 Eros* | *Near-Earth asteroid* | *Orbit (2 altitudes) and controlled collision* | NEAR Shoemaker/2000-01 | *16 km* |
| 951 Gaspra | Main belt asteroid | Flyby (primary: Jupiter) | Galileo/1991 | 12 km |
| 1P/Halley | Comet | Flyby (primary: --, Venus) | Gioto/1986, Vega/1986 | 11 km |
| 9P/Tempel | Comet | Flyby (primary: --) | Deep Impact/2005, NExT/2011 | 5 km |
| 5535 AnneFrank | Main belt asteroid | Flyby (primary: 81P/Wild) | Stardust/2002 | 5km |
| *67P/Churyumov-Gerasimenko* | *Comet* | *Orbit and land* | *Rosetta/2014* | *4 km* |
| 81P/Wild | Comet | Flyby (primary: --) and coma sample return | Stardust/2004 | 4 km |
| 2867 Šteins | Main belt asteroid | Flyby (primary: 67P/Churyumov-Gerasimenko | Rosetta/2008 | 2.65 km |
| 103P/Hartley | Comet | Flyby (primary: --) | EPOXI/2010 | 1.1 km |
| *25143 Itokawa* | *Near-Earth asteroid* | *Orbit and surface sample return* | *Hayabusa/2005* | *0.3 km* |

**Table 1.** Small-bodies observed to-date, ordered by diameter. Orbit missions are italicized. Flyby missions have their primary target identified ('--' means this small-body was the primary target). *actual diameter is unknown; this value has been estimated from absolute magnitude (Melita et al., 2010).

| | L4/Greek camp (Body ID and diameter (km)) | | L5/Trojan camp | |
|---|---|---|---|---|
| **Redder Trojans** | 588 Achilles | 135 | 884 Priamus | 101* |
| | 624 Hektor | 225 | 1172 Aneas | 143 |
| | 911 Agamemnon | 167 | 1867 Deiphobus | 123 |
| | 1143 Odysseus | 126 | 2207 Antenor | 85 |
| | 1583 Antilochus | 101 | 2223 Sarpedon | 94 |
| | 1868 Thersites | 81* | 2241 Alcathous | 115 |
| | 2260 Neoptolemus | 72 | 2363 Cebriones | 82 |
| | 2456 Palamedes | 92 | 2893 Peiroos | 87 |
| | 2759 Idomeneus | 61 | 3317 Paris | 116 |
| | 2797 Teucer | 111 | 5144 Achates | 92 |
| | 2920 Automedon | 111 | (12929) 1999 TZ1 | 81* |
| | 3063 Makhaon | 116 | (34746) 2001 QE91 | 84* |
| | 3540 Protesilaos | 94** | | |
| | 3564 Talthybius | 69 | | |
| | 3596 Meriones | 84* | | |
| | 3709 Polypoites | 99 | | |
| | (4035) 1986 WD | 69 | | |
| | 4063 Euforbo | 102 | | |
| | 4068 Menestheus | 63 | | |
| | 4833 Meges | 87 | | |
| | 4834 Thoas | 84 | | |
| | (4835) 1989 BQ | 65** | | |
| | 4902 Thessandrus | 70* | | |
| | 5027 Androgeos | 58 | | |
| | 5254 Ulysses | 78 | | |
| | 5264 Telephus | 73 | | |
| | 5283 Pyrrhus | 65 | | |
| | 5285 Krethon | 64* | | |
| | (7641) 1986 TT6 | 69 | | |
| | (9799) 1996 RJ | 65 | | |
| | (14690) 2000 AR25 | 51* | | |
| | (15436) 1998 VU30 | 86 | | |
| | (15440) 1998 WX4 | 66 | | |
| | (15527) 1999 YY2 | 49* | | |
| | (16974) 1998 WR21 | 55 | | |
| | (21595) 1998 WJ5 | 56* | | |
| | (21601) 1998 XO89 | 77* | | |
| | (21900) 1999 VQ10 | 64* | | |
| | (36267) 1999 XB211 | 42* | | |
| | (38050) 1998 VR38 | 77* | | |
| **Less-red Trojans** | 659 Nestor | 109 | 617 Patroclus | 141 |
| | 1437 Diomedes | 164 | 1173 Anchises | 126 |
| | 3548 Eurybates | 72 | 1208 Troilus | 103 |
| | 3793 Leonteus | 86 | 2895 Memnon | 81* |
| | 4060 Deipylos | 79 | 3451 Mentor | 140* |
| | 4138 Kalchas | 64* | (7352) 1994 CO | 92* |
| | (5025) 1986 TS6 | 57 | | |
| | 5244 Amphilochos | 56* | | |
| | (11395) 1998 XN77 | 65 | | |
| | (13385) 1998 XO79 | 59* | | |
| | (23135) 2000 AN146 | 73 * | | |

**Table 2.** List of Trojans, identified by Emery et al. (2011) as members of the "redder" or "less-red" spectral populations. Diameter data was taken from the NASA Small Bodies Database: *http://ssd.jpl.nasa.gov/sbdb_query.cgi* (July 31, 2012) when possible; otherwise they are *values reported in Emery et al. (2011) or **estimated diameters based on the reported absolute magnitude (*http://ssd.jpl.nasa.gov/sbdb_query.cgi* ) and assuming a geometric albedo of 0.05.

| | Science Goals | Measurement Objectives | Measurement Required | Instruments | Instrument Requirements | Data Products |
|---|---|---|---|---|---|---|
| Where/how did the Trojans form? | Constrain Origin of the Target Body | Measure volatile content | Fraction of volatiles to 1 m depth | ICING | At least 3 weeks mapping time for integration of signal; < 10° pointing accuracy (for all measurements) | Global map of H+ to 1 m depth, 40º $pixel^{-1}$ resolution |
| | | | Spectral signature of volatiles at surface to ~3 mm depth | CAVIAR | Spatial res.: IFOV = 250 μrad, 500 m $pixel^{-1}$ @ 1000 km, 150m $pixel^{-1}$ @ 300 km, 50m $pixel^{-1}$ @ 100km; Spectral res.: mapping ~ 20nm; pointing accuracy <25 μrad; spectral maps of 0.5-3.2 μm wavelengths | Global maps of volatile content at 1 m depth at 150 m $pixel^{-1}$; targeted maps at 50 m $pixel^{-1}$ |
| | | Determine rock-forming element content | Spectral signature of rock-forming elements to 1 m depth | ICING | See volatiles req. | See volatiles prod. – rock-forming elements |
| | | Identify mineral composition | Detection of ferrous solid solution series (+ other possible minerals) to 10% | CAVIAR | See volatiles req. -- spectral maps of 0.9-2.4 μm wavelengths | See volatiles prod. – for mineral composition |
| | Assess Compositional Heterogeneity | Identify geologic units | Discrimination between light/dark minerals (color differences, albedo changes of 2%) | WASABI | Full coverage @ 300 km orbit; 9 colors and 2 polarizations | Global color unit maps at 54 m $pixel^{-1}$; targeted areas at 18 m $pixel^{-1}$ |
| | | | Identification of compositional units | CAVIAR | See volatiles req. (in Origin of the Target Body) -- spectral maps of 0.5-2.5 μm wavelengths | See volatiles product (in Origin of the Target Body) – for compositional units |
| | | Measure water content | Hydrogen content to 1 m depth | ICING | See volatiles req. (in Origin of the Target Body) | See volatiles product (in Origin of the Target Body) |
| | Characterize Degree of Differentiation | Determine moment of inertia | Tracking of spacecraft orbit | RSCM | 1 mm $s^{-2}$ | 2-way Doppler readings |
| | | Measure gravity distribution | Geologic unit and topography data | WASABI-NACHO, CAVIAR | See geologic unit req.s (in Compositional Heterogeneity, Geologic History) | See geologic unit prod.s (in Compositional Heterogeneity, Geologic History) |
| | | | Tracking of spacecraft orbit | RSCM | 0.1 mm $s^{-2}$ | 2-way Doppler readings |
| **How have the Trojans evolved?** | Develop Geologic History | Identify geologic units | Distinguish surface structures > 25 m in size | NACHO | 8 m $pixel^{-1}$ resolution @ 300 km orbit; 12.5 µrads pointing accuracy (for all measurements) | Global imaging at 8 m $pixel^{-1}$; targeted imaging at 2.5 m $pixel^{-1}$ |
| | | Count and categorize craters | See geologic units req. | WASABI-NACHO | See geologic unit req. | Global crater counts down to 25 m diameter |
| | | Measure topography | Distinguish roughness > 30 m height | NACHO | 2.5m $pixel^{-1}$ @ 100 km orbit | Targeted stereo and high-phase angle images at 2.5 m $pixel^{-1}$ |
| | Determine Surface Properties and Weathering | Observe current state of communition | Fraction of volatiles at surface to ~3 mm depth | CAVIAR | See volatiles req. (in Origin of the Target Body) -- -- spectral maps of 0.5-3 µm wavelengths | See volatiles prod. |
| | | | Discrimination between cobbles and fine dust | NACHO | 2.5m $pixel^{-1}$ @ 100 km orbit taken at 5 phase angles: 0, 5, 10, 20, 40 | Estimates of surface particle size within targeted regions |
| | | Observe current state of compaction | Volatile and element distribution at surface to ~3 mm depth | CAVIAR | See volatiles req. (in Origin of the Target Body) – for volatile content and mineral composition | See volatiles prod. – for volatile content and mineral composition |
| | | | Distinguish surface geologic units | NACHO | See communition req. | See communition req. |
| | Detect and Identify Organic | Detect evidence of CHON/ C=C | Detect C=C, C=N, C-H bonds at 10% concentration | CAVIAR | See volatiles req. (in Origin of the Target Body) -- spectral maps of 1-4.5 µm range | See volatiles prod. (in Origin of the Target Body) – for organic |
| | Composition | | | | | detection |

**Table 3.** Science Traceability Matrix. Our proposed instrument suite: WASABI-NACHO is a dual-camera system, CAVIAR is a visible and near infrared mapping spectrometer, and ICING is a neutron and gamma-ray spectrometer. RSCM is the Radio Science Celestial Mechanics system (i.e., radio-tracking of the spacecraft).

<table>
<tr><th colspan="2">Instrument</th><th>Purpose</th><th>Estimated Mass (kg)</th><th colspan="2">Peak power (W)</th><th>Standby power (W)</th></tr>
<tr><td rowspan="2">Cameras</td><td>WASABI</td><td>Generate global maps of surface appearance/geology</td><td rowspan="2">8.8</td><td rowspan="2">20 (when operating together)</td><td>14</td><td rowspan="2">2</td></tr>
<tr><td>NACHO</td><td>Generate global visual maps at high-altitudes and targeted high-res images at low-altitude</td><td>10</td></tr>
<tr><td colspan="2">CAVIAR</td><td>Generate global maps of surface composition at high-altitudes and targeted high-res images at low-altitude</td><td>10.3</td><td colspan="2">22</td><td>2</td></tr>
<tr><td colspan="2">ICING</td><td>Detect and measure surface volatile content, generating global map</td><td>9.3</td><td colspan="2">15</td><td>3</td></tr>
</table>

**Table 4.** The proposed payload: WASABI-NACHO is a dual-camera system, CAVIAR is a visible and near infrared mapping spectrometer, and ICING is a neutron and gamma-ray spectrometer.

| Instrument | Purpose | Estimated Mass (kg) |
|---|---|---|
| Radar Ranger/Sounder | Map near-surface/subsurface features | 17 |
| Thermal/IR Imager | Determine albedo and thermal inertia | 11.9 |
| Laser Altimeter | Measure topographical profiles to determine geology and revolution | 6.1 |
| UV Spectrometer | Detect outgassing and identify heavy element oxides | 3.1 |
| Impactor (used only during flyby) | Expose subsurface elements | -- |

**Table 5.** Instruments that were considered, but descoped due to mission mass, power, and cost requirements.

| | Objectives | Operating Instruments | No. Observ. Returned | Spatial Res. | Data Volume (bits) | Science and Downlink Intervals |
|---|---|---|---|---|---|---|
| **Approach** (begins ~100 hrs before orbital insertion) | Calibrate instruments, Generate rough shape model and rotation characterization map | NACHO | -- | -- | -- | -- |
| **Survey Orbit** (1000 km, 120 hr period, 20 days @ 3pm) | Global imaging, Identify targets of interest, Begin generation of topography model, Finalize instrument calibrations | WASABI | 7 frames | 180 m $px^{-1}$ | 6.8e7 | 16 hours of science operations, followed by 8 hours of downlink (total data return: 7.5e9 bits) |
| | | NACHO | 57 frames | 25 m $px^{-1}$ | 5.5e8 | |
| | | CAVIAR | 1380 lines | 250 m $px^{-1}$ | 2.1e9 | |
| **High-altitude Mapping Orbit** (300 km, 24 hr period, 3 months @ 1pm) | Global imaging with all instruments, Identify targets of interest, Enhance topographic model | WASABI | 69 frames | 54 m $px^{-1}$ | 6.7e8 | Per orbit: 16 hours of science operations, 8 hours of downlink on the dark side (total data return: 3.4e10 bits) |
| | | NACHO | 623 frames | 7.5 m $px^{-1}$ | 6.0e9 | |
| | | CAVIAR | 15300 lines | 75 m $px^{-1}$ | 1.6e10 | |
| **Low-altitude Mapping Orbit** (100 km, 8 hr period, 6 months @ 2pm and 2 months @ noon) | Target previously identified regions-of-interest for high-resolution visual and spectral observations (> 30% coverage), Global imaging with WASABI and ICING | WASABI | 613 frames | 18 m $px^{-1}$ | 5.9e9 | 1 orbits (8 hours) of science operations, followed by 2 orbits (16 hours) of downlink (data total return: 1.8e11 bits) |
| | | NACHO | 4000 frames | 2.5 m $px^{-1}$ | 3.9e10 | |
| | | CAVIAR | 69000 lines | 25 m $px^{-1}$ | 7.0e10 | |
| | | ICING | Continuous | -- | 4.0e10 | |

**Table 6**. Rendezvous observation plan. We would include 3 altitudes to allow us to update navigation plans during the mission, similar to the Dawn mission (Rayman et al., 2006). Different orbit inclinations were chosen to include a range of illumination angles. The data volumes estimates assume compression (8:1 lossy compression for the cameras, as was used by MDIS (Hawkins at al.,

2007); 1.5 compression for CAVIAR, as was used by $M^3$ (Pieters et al., 2009)) and 15% overhead. WASABI-NACHO is a dual-camera system, CAVIAR is a visible and near infrared mapping spectrometer, and ICING is a neutron and gamma-ray spectrometer.

| Mission Cost Summary | Estimated Cost | Reserve | Total Cost |
|---|---|---|---|
| Phase A | 29.2 | 50% | 43.8 |
| Phase B | 93.3 | 50% | 139.7 |
| Phase C/D | 426.3 | 50% | 638.4 |
| Phase E | 149.3 | 23% | 183.4 |
| **Total PI-managed Project Cost** | **698.1** | **44%** | **1005.5** |

**Table 7.** Cost estimate, per mission phase, for the proposed TASTER mission. All values are in millions-FY15$.

| | ID | Incl.(°) | Eccen. | Diam. (km) | Abs. Mag. | Geometric Albedo |
|---|---|---|---|---|---|---|
| **L4 (Greek camp)** | 588 Achilles[■,▲] | 10.32 | 0.148 | 135 | 8.67 | 0.033 |
| | 624 Hektor[■,▲] | 18.179 | 0.023 | 225 | 7.49 | 0.025 |
| | 911 Agamemnon[†,■,▲] | 21.777 | 0.068 | 167 | 7.89 | 0.044 |
| | 1143 Odysseus[†,▲] | 3.138 | 0.092 | 126 | 7.93 | 0.075 |
| | 1437 Diomedes[■,▲] | 20.502 | 0.046 | 164 | 8.3 | 0.031 |
| | 2759 Idomeneus[▲] | 21.956 | 0.066 | 61 | 9.8 | 0.057 |
| | 2797 Teucer[▲] | 22.402 | 0.089 | 111 | 8.4 | 0.062 |
| | 3540 Protesilaos[■,▲] | 23.313 | 0.118 | | 9 | |
| | 3548 Eurybates[•] | 8.072 | 0.091 | 72 | 9.5 | 0.054 |
| | 3793 Leonteus[■,▲] | 20.923 | 0.091 | 86 | 8.8 | 0.072 |
| | (4035) 1986 WD[•] | 12.133 | 0.058 | 69 | 9.72 | 0.072 |
| | 4060 Deipylos[†,■,▲] | 16.149 | 0.156 | 79 | 8.9 | 0.078 |
| | 4063 Euforbo[■,▲] | 18.943 | 0.12 | 102 | 8.6 | 0.061 |
| | (4835) 1989 BQ[■,▲] | 19.572 | 0.252 | | 9.8 | |
| | (5025) 1986 TS6[•] | 11.02 | 0.761 | 57.83 | 9.8 | 0.064 |
| | 5254 Ulysses■,▲ | 24.194 | 0.122 | 78.34 | 8.8 | 0.087 |
| | (6545) 1986 TR6● | 11.99 | 0.053 | | 10 | |
| **L5 (Trojan camp)** | 617 Patroclus■,▲ | 22.053 | 0.14 | 141 | 8.19 | 0.047 |
| | 1172 Aneas■,▲ | 16.675 | 0.106 | 143 | 8.33 | 0.04 |
| | 1173 Anchises● | 6.915 | 0.138 | 126 | 8.99 | 0.031 |
| | 1867 Deiphobus■,▲ | 26.91 | 0.044 | 123 | 8.61 | 0.042 |
| | 2223 Sarpedon● | 15.968 | 0.017 | 94 | 9.25 | 0.034 |
| | 2357 Phereclos● | 2.669 | 0.045 | 95 | 8.86 | 0.052 |
| | 3451 Mentor‡ | 24.68 | 0.073 | 140 | 8.1 | |
| | (3708) 1974 FV1■ | 13.37 | 0.16 | 80 | 9.3 | 0.053 |
| | 4348 Poulydamas■ | 7.96 | 0.099 | | 9.2 | |
| | 5144 Achates■,▲ | 8.902 | 0.272 | 92 | 8.9 | 0.058 |
| | 5511 Cloanthus● | 11.176 | 0.118 | 55 | 10.43 | |

**Table 8.** Possible rendezvous targets for a Trojan tour mission, based on identification as a viable mission target in a prior Trojan mission design study: †Brown et al., 2010 or ‡Simanjuntak et al. (2010), or as a target in ground-based telescope observations: ■Emery and Brown (2003), ▲Emery and Brown (2004), or ●Fornasier et al. (2007). As we are primarily interested in identifying rendezvous targets (versus flyby targets), we list only larger bodies (i.e., the Trojan needs to have absolute magnitude < 10 or an estimated diameter of > 50 km; Lamy et al. (2012) was also checked, but all objects in that proposed flyby tour had absolute magnitude > 10). Diameter information is from ▲Emery and Brown (2004), ●Fornasier et al. (2007). Orbit and albedo information comes from those papers and the NASA Small Bodies Database: *http://ssd.jpl.nasa.gov/sbdb_query.cgi*.

**Figures**

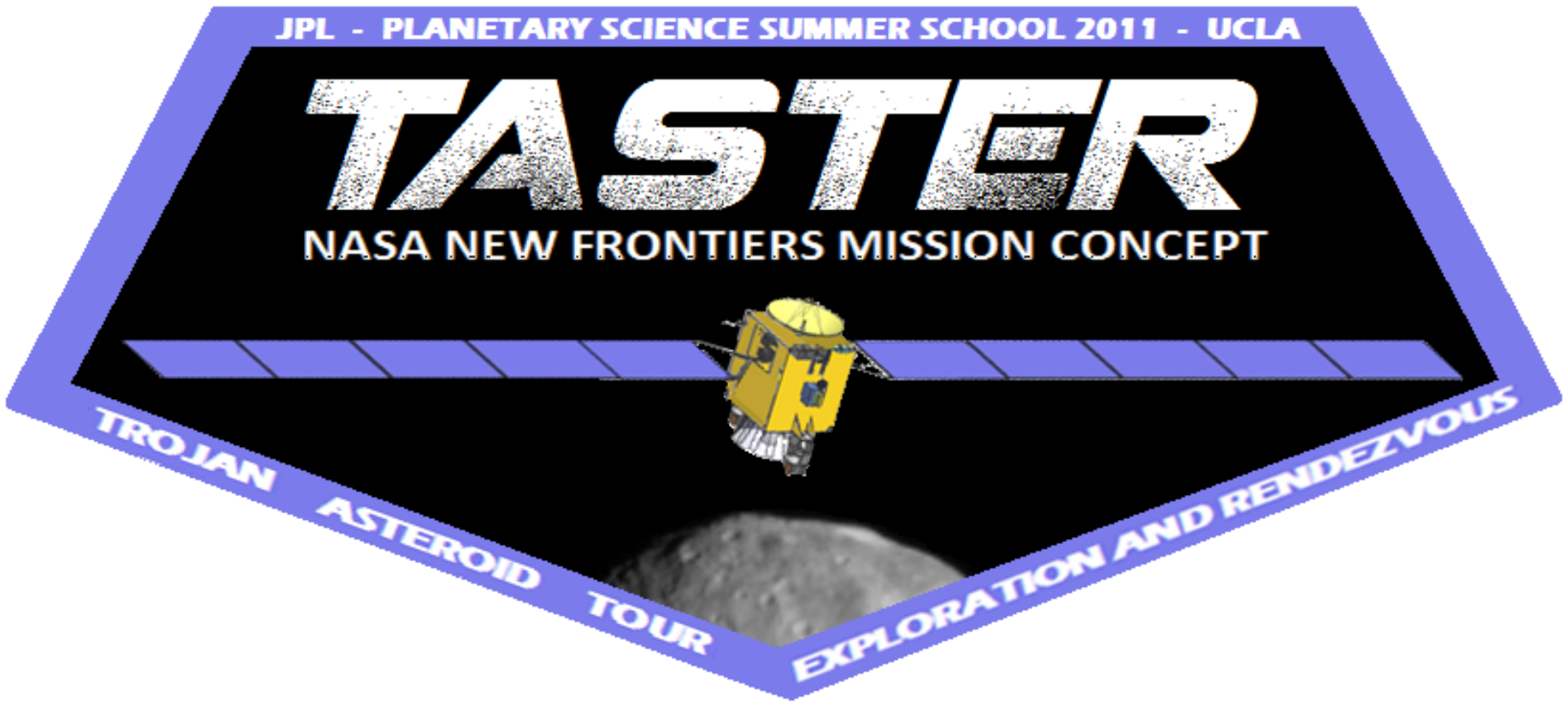

**Figure 1.** Our mission logo, of our spacecraft orbiting a small body, was designed by participant Ricardo Diaz-Silva. UCLA was highlighted as that was the home institution of our PI, Kunio Sayanagi.

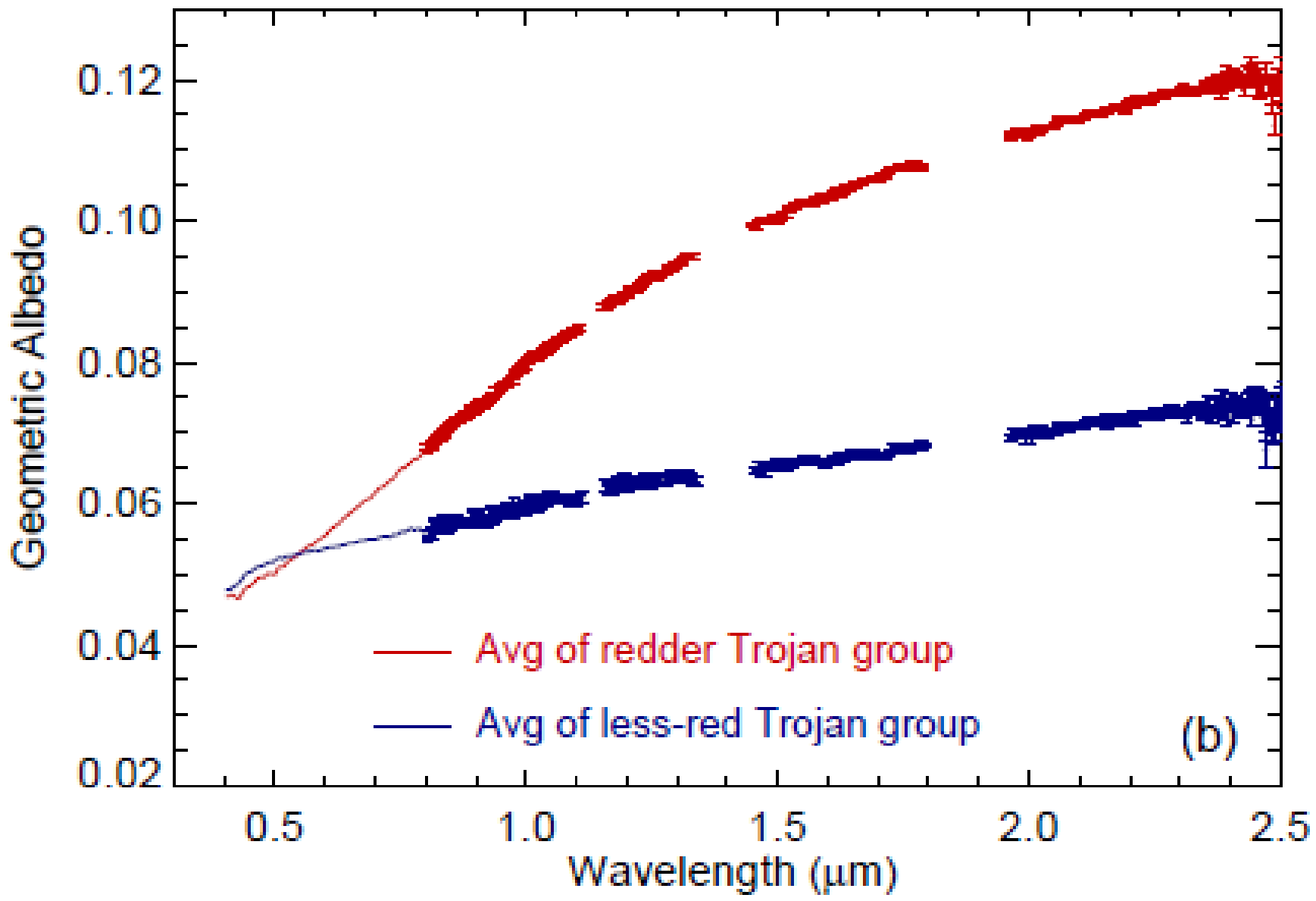

**Figure 2.** Averaged spectra (normalized at the albedo 0.0503) for the redder (upper curve) and less-red Trojan groups, which clearly show the difference in spectral slope. The figure is from Emery et al. (2011, Figure 6).

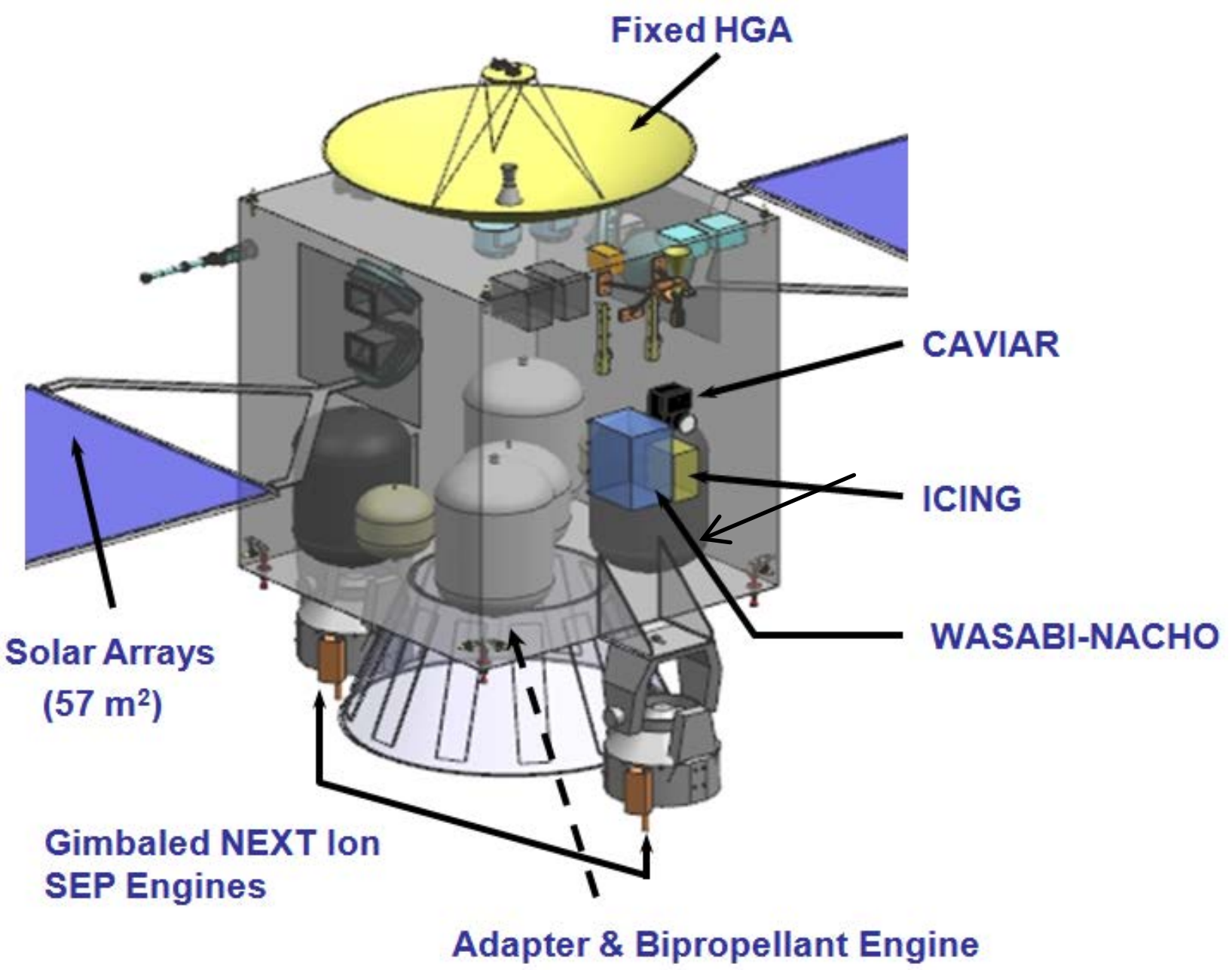

**Figure 3.** A CAD model of the TASTER spacecraft structure. The HGA dish is on top of the 2 m-cube bus. Extending from the sides are the articulating arms for the solar arrays (trimmed). A partially transparent view shows the propellant tanks located inside the spacecraft. The HiPAT engine is located in the center, between the two gimbaled NEXT engines and partially obstructed from view by the spacecraft's adapter ring.

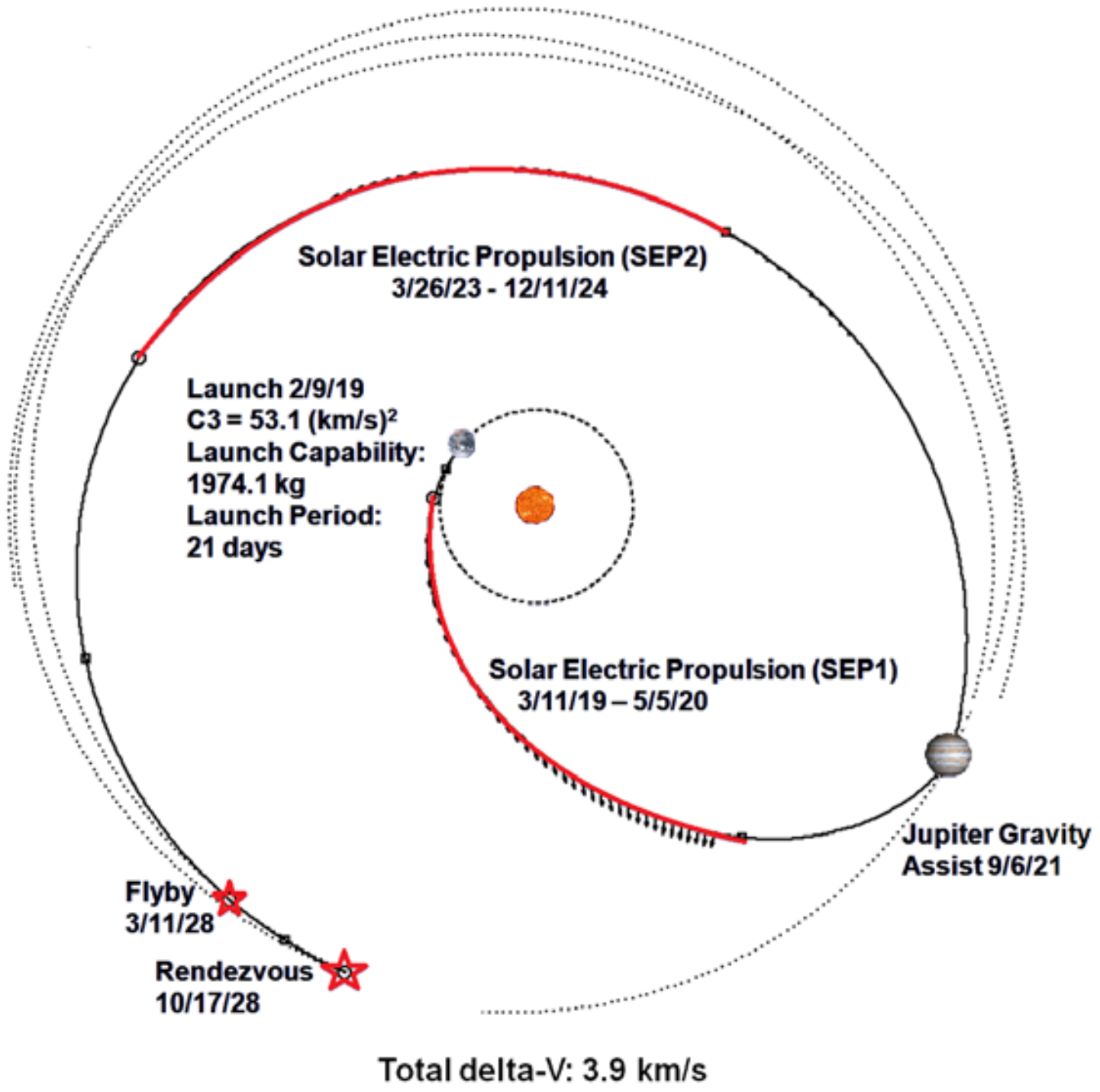

**Figure 4.** The proposed spacecraft trajectory. The total delta-V required to reach our targets is 3.914 km $s^{-1}$ and is supplied by a combination of chemical and ion propulsion (the burns for the latter are demarked by the thicker line in the trajectory). The Jupiter Gravity Assist is used to make a plane change.

## Acknowledgements

This document was created by students as an educational activity at the Jet Propulsion Laboratory, California Institute of Technology, and does not represent an actual mission.  Government sponsorship acknowledged. We thank everyone involved with the NASA-JPL Planetary Science Summer School and Team X for enriching our experience, with special thanks to Charles Budney for serving as our mentor and Leslie Lowes and Trisha Wheeler for their logistics assistance. We also thank our review board for their insight and advice: Mark Adler, Bruce Banerdt, Rosaly Lopes, Edward Miller, and Adam Steltzner. Finally, we thank our reviewers, Andrew Rivkin and Josh Emery, for their insightful and constructive comments. Diniega was supported by an appointment to the NASA Postdoctoral Program, administered by Oak Ridge Associated Universities, at the California Institute of Technology Jet Propulsion Laboratory under a contract with NASA.

## References

Bender, D.F., Penzo, P.A., 1995. Multiple encounter missions with the Trojan and Main Belt Asteroids. AAS/AIAA Astrodynamics Specialist Conference, Halifax, Nova Scotia, Canada. Report AAS 95-0134.

Binzel, R.P., Sauter, L.M., 1992. Trojan, Hilda, and Cybele asteroids: New lightcurve observations and analysis. Icarus 95(2), 222-238.

Bonello, G., Bibring, J-P., Poulet, F., Gendrin, A., Gondet, B., Langevin, Y., Fonti, S., 2004. Visible and infrared spectroscopy of minerals and mixtures with the OMEGA/MARS-EXPRESS instrument. Planet. Space Sci. 52, 133-140. doi:10.1016/j.pss.2003.08.014.

Bonfiglio, E.P., Oh, D., Yen, C., 2005. Analysis of Chemical, REP, and SEP missions to the Trojan asteroids. AAS/AIAA Astrodynamics Specialist Conference, Lake Tahoe, CA. Report AAS 05-396.

Bottke, W.F., Durda, D.D., Nesvorný, D., Jedicke, R., Morbidelli, A., Vokrouhlický, D., Levison, H., 2005. The fossilized size distribution of the main asteroid belt. Icarus 175, 111-140.

Brown, M. and 28 coauthors, 2010. Trojan Tour Decadal Study. NASA SDO-12348.

Davis, D.R.; Weidenschilling, S.J.; Farinella, P.; Paolicchi, P.; Binzel, R.P., 1989. Asteroid collisional history – Effects on sizes and spins. IN: Asteroids II; Proceedings of the Conference, Tucson, AZ, Mar. 8-11, 1988 (A90-27001 10-91). Tucson, AZ, University of Arizona Press, 805-826.

De Luise, F. Dotto, E., Fornasier, S., Barucci, M.A., Pinilla-Alonso, N., Perna, D., Marzari, F., 2010. A peculiar family of Jupiter Trojans: The Eurybates. Icarus, 209, 586-590. doi:10.1016/j.icarus.2010.04.024.

Department of Energy (DOE), 2010. Startup Plan for Plutonium238 Production for Radioisotope Power Systems, Report to Congress.

Dotto, E., Emery, J.P., Barucci, M.A., Morbidelli, A., Cruikshank, D.P., 2008. De Troianis: The Trojans in the planetary system. IN: The Solar System beyond Neptune, M. A. Barucci, H. Boehnhardt, D. P. Cruikshank, and A. Morbidelli (*eds.*), University of Arizona Press, 383-395.

Emery, J.P., Brown, R.H., 2003. Constraints on the surface composition of Trojan asteroids from near-infrared (0.8-4.0 μm) spectroscopy. Icarus 164, 104-121. DOI 10.1016/S0019-1035(03)00143-X.

Emerry, J.P. and Brown, R.H., 2004. The surface composition of Trojan asteroids: constraints set by scattering theory. Icarus 170, 131-152. doi:10.1016/j.icarus.2004.02.004.

Emery, J.P., Cruikshank, D.P., Van Cleve, J., 2006. Thermal emission spectroscopy (5.2–38 μm) of three Trojan asteroids with the Spitzer Space Telescope: Detection of fine-grained silicates. Icarus 182, 496-512. doi:10.1016/j.icarus.2006.01.011.

Emery, J. P., Burr, D. M., Cruikshank, D. P., 2011. Near-infrared Spectroscopy of Trojan Asteroids: Evidence for Two Compositional Groups. Astronom. J. 141(1), article id. 25.

Fernandez, Y.R., Sheppard, S.S., Jewitt, D.C., 2003. The albedo distribution of Jovian Trojan asteroids. Astronom. J. 126, 1563-1574.

Fornasier, S., Dotto, E., Hainaut, O., Marzari, F., Boehnhardt, H., De Luise, F., Barucci, M.A., 2007. Visible spectroscopic and photometric survey of Jupiter Trojans: Final results on dynamical families. Icarus 190, 622-642. doi:10.1016/j.icarus.2007.03.033.

Gomes, R., Levison, H.F., Tsiganis, K., Morbidelli, A., 2005. Origin of the cataclysmic Late Heavy Bombardment period of the terrestrial planets. Nature 435, 466-469. DOI 10.1038/nature03676.

Gradie, J., Veverka, J., 1980. The composition of the Trojan asteroids. Nature 283(5750), 840-842.

Grammier, R.S., 2009. A look inside the Juno Mission to Jupiter. Proceedings of 2009 IEEE Aerospace conference, 1-10. doi: 10.1109/AERO.2009.4839326.

Hawkins, S.E., Boldt, J.D., Darlington, E.H., Espiritu, R., Gold, R.E., Gotwols, B., Grey, M. P., Hash, C.D., Hayes, J.R., Jaskulek, S.E., Kardian, C.J., Keller, M.R., Malaret, E.R., Murchie, S.L., Murphy, P.K., Peacock, K., Prockter, L.M., Reiter, R.A., Robinson, M.S., Schaefer, E.D., Shelton, R.G., Sterner, R.E., Taylor, H.W., Watters, T.R., Williams, B.D., 2007. The Mercury Dual Imaging System on the MESSENGER Spacecraft. Space Sci Rev 131, 247-338. DOI 10.1007/s11214-007-9266-3.

Hevey, P., Sanders, I.S., 2006. A model for planetesimal meltdown by $^{26}$Al and its implications for meteoritic parent bodies. Meteor. Planet. Sci. 41, 95-106.

Lacerda, P., Jewitt, D.C., 2006. Densities of Solar System objects from their rotational light curves. Astronom. J. 133, 1393-1408.

Lamy, P., Vernazza, P.; Poncy, J.; Martinot, V.; Hinglais, E.; Canalias, E.; Bell, J.; Cruikshank, D.; Groussin, O.; Helbert, J.; Marzari, F.; Morbidelli, A.; Rosenblatt, P.; Sierks, H., 2012. Trojans' odyssey: Unveiling the early history of the Solar System. Exp. Astron., published online. doi 10.1007/s10686-011-9253-2.

Margot, J.L., Brown, M.E., 2003.A low-density M-type asteroid in the Main Belt. Science 300, 1939-1942. doi:10.1126/science.1085844.

Marzari, F., Farinella, P., Davis, D. R., Scholl, H., Campo Bagatin, A., 1997. Collisional evolution of Trojan asteroids. Icarus 125(1), 39-49. DOI 10.1006/icar.1996.5597.

Marzari, F., Scholl, H., 1998. Capture of Trojans by a growing proto-Jupiter. Icarus 131, 41-51.

Melita, M.D., Licandro, J., Jones, D.C., Williams, I.P., 2008. Physical properties and orbital stability of the Trojan asteroids. Icarus 195, 686-697.

Melita, M.D., Duffard, R., Williams, I.P., Jones, D.C., Licandro, J., Ortiz, J.L., 2010. Lightcurves of 6 Jupiter Trojan asteroids. Planet. Space Sci. 58(7-8), 1035-1039. http://dx.doi.org/10.1016/j.pss.2010.03.009

Morbidelli, A., Chambers, J., Lunine, J.I., Petit, J.M., Robert, F., Valsecchi, G.B., Cyr, K.E., 2000. Source regions and timescales for the delivery of water to the Earth. Meteor. Planet. Sci. 35, 1309-1320.

Morbidelli, A., Levison, H.F., Tsiganis, K., Gomes, R., 2005. Chaotic capture of Jupiter's Trojan asteroids in the early Solar System. Nature 435, 462-465. doi: 10.1038/nature03540.
Mueller, M., Marchis, F., Emery, J.P., Harris, A.W., Mottola, S., Hestroffer, D., Berthier, J., di Martino, M., 2010. Eclipsing binary Trojan asteroid, Patroclus: Thermal inertia from Spitzer observations. Icarus 205, 505-515. doi:10.1016/j.icarus.2009.07.043.
National Research Council (NRC), 2011. Vision and Voyages for Planetary Science in the Decade 2013-2022, The National Academies Press, Washington, D.C.
National Aeronautics and Space Administration (NASA), 2009. Announcement of Opportunity: New Frontiers Program. NASA NNH09ZDA007O.
Patterson, M.J., Benson, S.W., 2007. NEXT Ion Propulsion System Development Status and Capabilities. NASA Science Technology Conference, Paper D10P3.
Perozzi, E., Rossi, A., Valsecchi, G.B., 2001. Basic targeting strategies for rendezvous and flyby missions to the near-Earth asteroids. Planet. Space Sci. 29, 3-22.
Pieters, C.M., Boardman, J., Buratti, B., Chatterjee, A., Clark, R., Glavich, T., Green, R., Head, J., Isaacson, P., Malaret, E., McCord, T., Mustard, J., Petro, N., Runyon, C., Staid, M., Sunshine, J., Taylor, L., Tompkins, S., Varanasi, P., Whit, M., 2009. The Moon Mineralogy Mapper ($M^3$) on Chandrayaan-1. Current Science 96(4), 500-505.
Prettyman, T.H., Feldman, W.C., Ameduri, F.P., Barraclough, B.L., Cascio, E.W., Fuller, K.R., Funsten, H.O., Lawrence, D.J., McKinney, G.W., Russell, C.T., Soldner, S.A., Storms, S.A., Szeles, C., Tokar, R.L., 2003. Gamma-Ray and Neutron Spectrometer for the Dawn Mission to 1 Ceres and 4 Vesta. IEEE Transactions on Nuclear Science 50(4), 1190-1197.
Prettyman, T.H., Hagerty, J.J., Elphic, R.C., Feldman, W.C., Lawrence, D.J., McKinney, G.W., Vaniman, D.T., 2006. Elemental composition of the lunar surface: Analysis of gamma ray spectroscopy data from Lunar Prospector. J. Geophys. Res. 111, E12007. doi: 10.1029/2005JE002656.
Prettyman, T.H., Feldman, W.C., McSween Jr., H.Y., Dingler, R.D., Enemark, D.C., Patrick, D.E., Storms, S.A., Hendricks, J.S., Morgenthaler, J.P., Pitman, K.M., Reedy, R.C., 2011. Dawn's Gamma Ray and Neutron Detector. Space. Sci. Rev. 163, 371-459. doi: 10.1007/s11214-011-9862-0.
Rayman, M.D.; Fraschetti, T.C.; Raymond, C.A.; Russell, C.T., 2006. Dawn: A mission in development for exploration of main belt asteroids Vesta and Ceres. Acta Astronautica 58(11), 605-616.
Rayman, M.D., Fraschetti, T.C., Raymond, C.A., Russel, C.T., 2007. Coupling of system resource margins through the use of electric propulsion: Implications in preparing for the Dawn mission to Ceres and Vesta. Acta Astronaut. 60, 930-938.
Rivkin, A., Emery, J., Barucci, A., Bell, J.F., Bottke, W.F., Dotto, E., Gold, R., Lisse, C., Licandro, J., Prockter, L., Hibbits, C., Paul, M., Springmann, A., Yang, B., 2010. The Trojan asteroids: Keys to many locks. Decadal White Paper.
Russell, C. T.; Capaccioni, F.; Coradini, A.; de Sanctis, M. C.; Feldman, W. C.; Jaumann, R.; Keller, H. U.; McCord, T. B.; McFadden, L. A.; Mottola, S.; Pieters, C. M.; Prettyman, T. H.; Raymond, C. A.; Sykes, M. V.; Smith, D. E.; Zuber, M. T., 2007. Dawn mission to Vesta and Ceres: Symbiosis between terrestrial observations and robotic exploration. Earth, Moon, and Planets 101(1-2), 65-91.
Schulz, R., Benkhoff, J., 2006. BepiColombo: Payload and mission updates. Advanc. Space Res. 38, 572-577.

Simanjuntak, T., Nakamiya, N., Kawakatsu, Y., 2010. Mission Design for Jupiter Trojans Rendezvous Mission. AAS/AIAA Space Flight Mechanics Meeting, San Diego, CA. Report AAS 10-0246.
Tsiganis, K., Gomes, R., Morbidelli, A., Levison, H.F., 2005. Origin of the orbital architecture of the giant planets of the Solar System. Nature 435, 459-461. DOI 10.1038/nature03539.
Wall, S., 2000. Use of concurrent engineering in space mission design, Proceedings of EuSEC 2000, Munich, Germany.
Wall, S.D., 2004. Model-based engineering design for space missions. Proceedings of IEEE Aerospace Conference, 6, 3907-3915. doi: 10.1109/AERO.2004.1368208.
Yang, B., Jewitt, D., 2007. Spectroscopic search for water ice on Jovian Trojan asteroids. Astronom. J. 134, 223-228.
Yoshida, R., Nakamura, T., 2005. Size Distribution of faint Jovian L4 Trojan asteroids. Astronom. J. 130, 2900-2911.